\documentstyle[aps,twocolumn,epsfig]{revtex}
\begin{document}
\title{Detection of the BCS transition of a trapped Fermi Gas}
\author{Georg M.\ Bruun and Charles W.\ Clark}
\address{Electron and Optical Physics Division, 
National Institute of Standards and Technology, US Department of Commerce,
Gaithersburg, 
Maryland 20899-8410}
\maketitle
\begin{abstract}
We investigate theoretically the properties of a 
trapped gas of fermionic atoms in both the normal 
and the superfluid phases. Our analysis, which 
accounts for the shell structure 
of the normal phase spectrum, 
identifies two observables which 
are sensitive to the presence of the 
superfluid: the response of the gas to a modulation 
of the trapping frequency, and the heat capacity. 
Our results are discussed in the context 
of experiments on trapped Fermi gases.
\end{abstract}
The observation of Bose-Einstein condensation in 
several atomic systems~\cite{BEC} has recently
sparked increasing interest in trapped fermionic atoms. 
These systems offer the prospect of a Bardeen-Cooper-Schrieffer
(BCS) transition to a superfluid phase at low 
temperatures $T < T_c$. 
By trapping the atoms in two hyperfine states, the 
phase transition temperature $T_c$ 
should be experimentally accessible\cite{StoofBCS,Houbiers},
and several experimental groups are presently working 
to achieve this transition~\cite{JinBoulder,Mewescon}.
However, as only a few percent of the atoms 
are likely to participate in Cooper pairing~\cite{Tinkham},
it is not obvious how the transition could be 
observed in these dilute systems. Recently, it 
has been proposed that the propagation and 
scattering of light should be significantly 
altered by the presense of Cooper pairs~\cite{Ruostekoski,Wang}. 
Since the quasiparticles (QP) with energies near the Fermi
chemical potential, $\mu_{\rm F}$,  are those most affected 
by the Cooper pairing, candidate observables for the
detection of the BCS transition should be sought from
phenomena sensitive to this low-energy
region of the QP spectrum.

In this paper, 
we consider two such observables: the response
of the gas to a ``shaking'' of the trap,
as first suggested by Baranov~\cite{Baranov}; 
and the heat capacity. For low $T$, both of these
observables are dominated by contributions from
the low-energy spectrum. 
By presenting a complete calculation of the properties of the trapped gas 
in both the normal and superfluid phases, 
which accounts exactly for the quantization of 
the single-particle energy levels, we are able to predict 
if these two observables are suitable to detect the
presence of Cooper pairing. 
Our analysis should have direct relevance 
to the ongoing experiments
on trapped Fermi gases.

We consider a gas of fermionic atoms of mass $m$, confined by a potential
 $U_0({\mathbf{r}})$, with an equal number of atoms $N_\sigma$ in each
of two hyperfine states, $|\sigma=\pm\rangle$. 
Two fermions in the same internal state $\sigma$ must
have odd relative orbital angular momentum
(minimally $p$-wave), and at low temperatures the centrifugal
barrier suppresses their mutual interaction~\cite{DeMarco}.
Thus, we suppose the interaction to be effective only between atoms 
in different hyperfine states and to be dominated by the $s$-wave contribution.  
As the interplay between the discrete nature of the normal phase spectrum and the 
Cooper pairing is crucial for the quantities considered in this paper, we need a theory
 which can describe the interaction and Cooper pairing 
of atoms residing in different discrete trap levels. This precludes the
use of a simple Thomas-Fermi treatment \cite{Butts}. A theory appropriate for the present paper 
 has recently been presented\cite{BruunBCS}. 
It uses a zero-range pseudopotential~\cite{Huang} to 
model the interaction between atoms in two 
different hyperfine states; this is appropriate when the scattering
length for binary atomic collisions, $a$,
has a larger magnitude than the effective 
range of the interaction, $r_e$, and when
 $k_{\rm F}|a|\ll 1$, where $k_{\rm F}=\sqrt{2m\mu_{\rm F}/\hbar}$
is the Fermi wavevector. The generalized mean field theory derived from
this approach yields the eigenvalue problem~\cite{BruunBCS}:
\begin{eqnarray} 
E_\eta u_\eta({\mathbf{r}}) &=& [{\cal H}_0 + W({\mathbf{r}})]
u_\eta({\mathbf{r}}) + 
\Delta({\mathbf{r}}) v_\eta({\mathbf{r}})\nonumber \\
E_\eta v_\eta({\mathbf{r}}) &=& -[{\cal H}_0 + W({\mathbf{r}})]
v_\eta({\mathbf{r}}) + 
\Delta({\mathbf{r}}) u_\eta({\mathbf{r}})\label{BdG}.
\end{eqnarray}
Here ${\mathcal{H}}_0= -\frac{\hbar^2}{2m}\nabla^2+U_0({\mathbf{r}})-\mu_{\rm F}$ 
is the single-particle Hamiltonian; $W({\mathbf{r}})\equiv 
g\langle\hat{\psi}_{\sigma}^{\dagger}
({\mathbf{r}})\hat{\psi}_{\sigma}({\mathbf{r}})\rangle$
is the Hartree potential, where $\hat{\psi}_{\sigma}({\mathbf{r}})$ is the
atom field operator for component $\sigma$ at position {\bf r}, which obeys the usual fermion
anticommutation relations.  The coupling constant  is $g=4\pi a\hbar^2/m$ and the 
pairing field, $\Delta({\mathbf{R}})$, is defined
by \begin{equation} \label{Gapdef}
\Delta({\mathbf{R}})\equiv -g\lim_{r\rightarrow 0}\partial_r
[r\langle\hat{\psi}_{+}({\mathbf{R}}+\frac{{\mathbf{r}}}{2})
\hat{\psi}_{-}({\mathbf{R}}-\frac{{\mathbf{r}}}{2})\rangle].
\end{equation}
Our definition of the pairing field differs from
that often employed in weak-coupling BCS theory~\cite{deGennes}. The main 
advantage of the definition given above is that  it eliminates the
ultraviolet divergence present in the usual weak-coupling theory\cite{BruunBCS}.
The elementary quasiparticles (QPs) with excitation energies $E_\eta$ are described
by the Bogoliubov wave functions $u_\eta({\mathbf{r}})$ and $v_\eta({\mathbf{r}})$.

We solve the Bogoliubov-de Gennes (BdG) equations (\ref{BdG})
for the case of an isotropic harmonic potential,
$U_0(r)=m\omega^2r^2/2$, using a self-consistent numerical procedure 
outlined elsewhere~\cite{BruunBCS}. 
In the absence of the pairing field, the QPs exhibit a discrete spectrum
of energies $E_\eta$, with the index $\eta$ designating a triple 
of quantum numbers $(n, l,m)$, where $l,m$ are the usual angular
momentum quantum numbers and $n$ is an index of
radial excitation. In the presence of the pairing field, the self-consistent solution 
to the BdG equations with the lowest free energy is spherically symmetric. Thus, the 
 Bogoliubov wavefunctions are given by $u_\eta({\mathbf{r}}) = r^{-1}u_{nl}(r)Y_{lm}(\theta,\phi)$ 
and $v_\eta({\mathbf{r}}) = r^{-1}v_{nl}(r)Y_{lm}(\theta,\phi)$, 
where the $Y_{lm}$ are the usual spherical harmonics,
and with $n,l,m$ being implicitly indexed by $\eta$. The paring field,  which is a
scalar operator under rotations, couples a normal phase QP with $(n,l,m)$ 
to one with $(n',l,-m)$.   Due to the spherical symmetry, in taking sums over states
needed to obtain the results of this paper, we can replace
sums over $m$ by factors of $(2l+1)$.

We now calculate the response of the gas to 
a harmonic time-dependent perturbation of the
trapping potential, $\Delta{\mathcal{H}}(t)$, of the form
\begin{equation}
\Delta{\mathcal{H}}(t)=
\lambda \sin(\tilde{\omega} t)\sum_\sigma\int\!d^3r\frac{1}{2}m\omega^2
r^2\psi_\sigma^\dagger ({\mathbf{r}})\psi_\sigma({\mathbf{r}}),
\label{perturb}
\end{equation}
where $\lambda$ is a small parameter. We  expand the
field operators in terms of the Bogoliubov wave functions and the QP operators in the
 usual way~\cite{deGennes}, and by  applying  
Fermi's golden rule to  obtain the linear 
response $R(\tilde{\omega})$ of the gas to 
the perturbation, Eq.(\ref{perturb}), we obtain:
\begin{eqnarray}\label{Abs}
R(\tilde{\omega})&\propto& 2\sum_{n>n',l}(2l+1)|
\int_{0}^{\infty}dr
(u_{nl} u_{n'l}-v_{nl} v_{n'l})r^2|^2\nonumber\\&\times&
(f_{n'l}-f_{nl})\delta(\hbar\tilde{\omega}+E_{n'l}-
E_{nl})+\nonumber\\&&
\sum_{n,n',l}(2l+1)|\int_{0}^{\infty} dr
(u_{nl} v_{n'l}+v_{nl} u_{n'l})r^2|^2\nonumber\\
 &\times&(1-f_{nl}-f_{n'l})\delta(\hbar\tilde{\omega}-E_{nl}-E_{n'l}),
\end{eqnarray}
where $f_{nl}=(\exp\beta E_{nl}+1)^{-1}$, $\beta=1/k_{\rm B}T$, and $k_{\rm B}$ is
Boltzmann's factor. The physical interpretation of the 
two terms in Eq.(\ref{Abs}) is straightforward:
The first term describes the excitation of a QP 
due to the perturbation, whereas 
the second term describes the creation of two QPs. This latter process
does not violate particle conservation,
since the QPs in general are mixtures of 
real particles and holes. The 
response of the gas should be observable 
as density fluctuations of the trapped gas.
As we have assumed a spherical symmetric perturbation, 
the transitions all 
have $\Delta l=0$. A 
generalization to perturbations with 
arbitrary angular momentum $l$ is straightforward.
In the non-interacting limit, Eq.(\ref{Abs}) 
reduces to a sum of delta functions 
$\delta(\tilde{\omega}-2n\omega)$ with $n=0,1,2\ldots$.

\begin{figure}
\centering
\epsfig{file=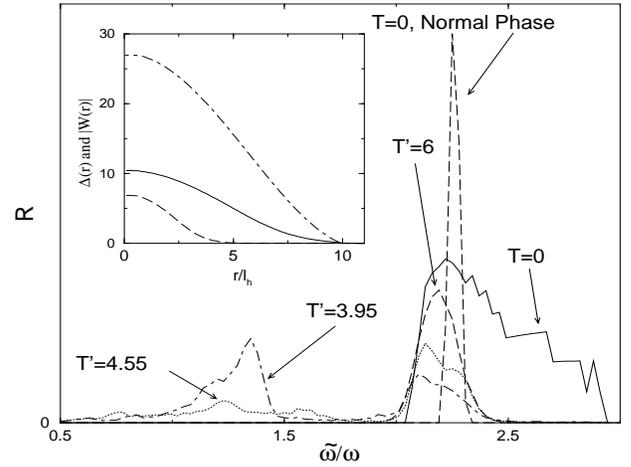,height=0.45\textwidth,width=0.27\textheight,
angle=-90}
\caption{The response $R$ of the gas as a function of the modulation
frequency 
$\tilde{\omega}$ for various temperatures. The inset shows the 
Hartree field $|W(r)|$ (dot-dashed) and $\Delta(r)$ for $T'=0$ (solid) 
and $T'=4.55$ (dashed) in 
units of $\hbar\omega$.}
\label{Shakefig}
\end{figure}

We now solve the BdG equations self-consistently and 
then calculate the response of the gas to a ``shaking'' 
of the trap from Eq.(\ref{Abs}). In Fig.\ref{Shakefig}, 
we show a typical plot of the response 
$R(\tilde{\omega})$ for various values of
$T'\equiv k_{\rm B}T/\hbar\omega$. In this example, 
we have chosen the parameters  $g/(\hbar\omega l_h^3)=-0.8$ and
 $\mu_{\rm F}=51.5\hbar\omega$, where $l_h=(\hbar/m\omega)^{1/2}$ is the 
characteristic length of the ground-state
harmonic oscillator wavefunction.
With the value of $a=-2160a_0$, appropriate to 
$^6$Li~\cite{Abraham}, these parameters correspond to
$N_\sigma\sim3.8\times10^4$ atoms of each spin state
in a trap with frequency $\nu=\omega/2\pi\simeq 520$Hz;
a value of $T_c \simeq 5.6\hbar\omega/k_{\rm B} = 140 {\rm nK}$ 
for the transition temperature is obtained by linearizing
Eq.(\ref{BdG})~\cite{BruunBCS}.  Fig.\ref{Shakefig} 
shows the response for $T'=0, 3.95$ and $4.55$, 
where the gas is in the superfluid phase, and 
for $T'=6>T_c$, where the gas is in the normal phase. 
For comparison, we also plot the $T=0$ response,
assuming the gas is in the normal phase. 
Each delta function in Eq.(\ref{Abs})
representing a $t\rightarrow\infty$
resonance, is smoothed out to a frequency 
range of $\sim\omega/10$ to model the 
finite frequency resolution of the
appropriate experiment. 

\begin{figure}
\centering
\epsfig{file=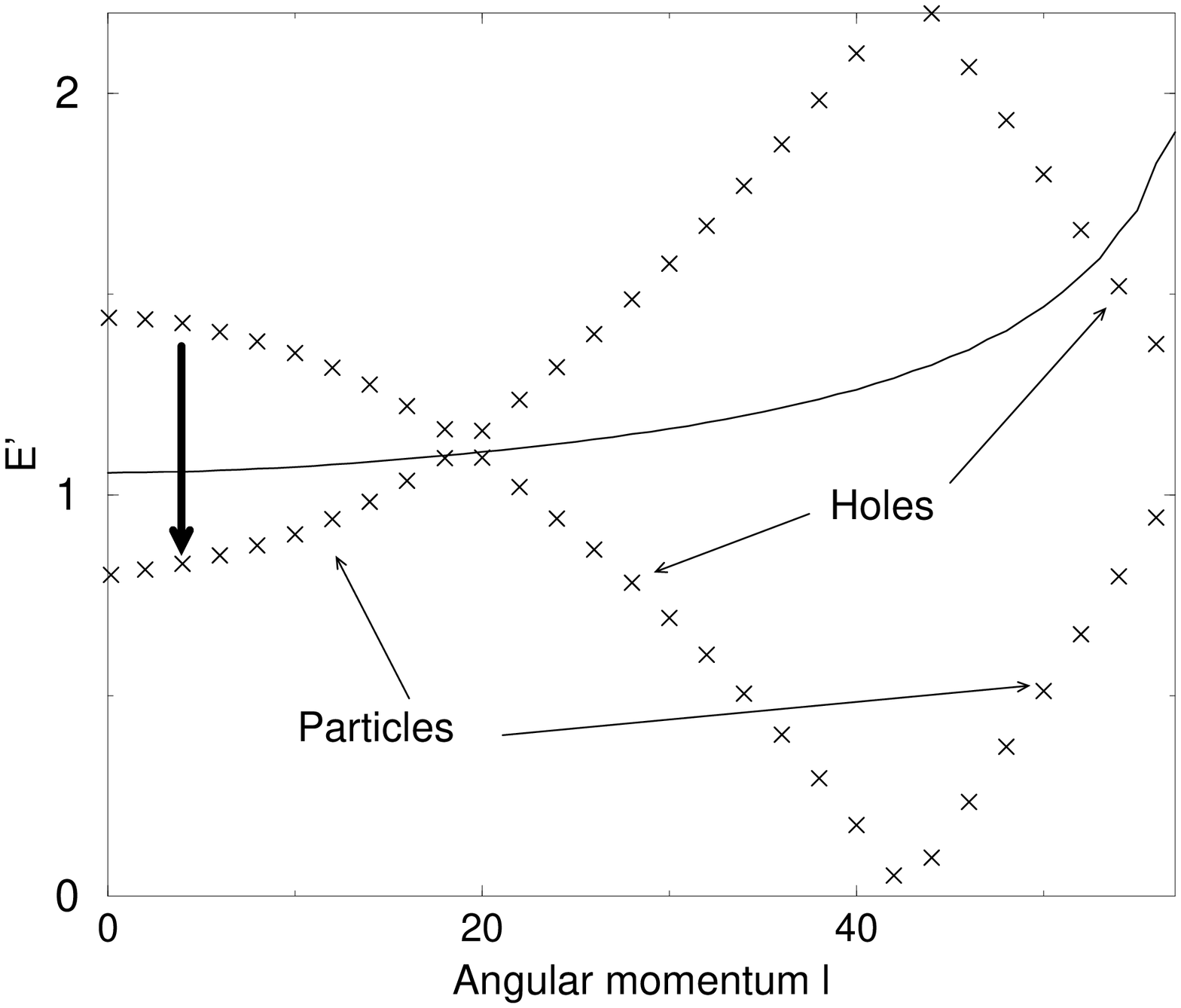,height=0.45\textwidth,width=0.24\textheight,
angle=-90}
\caption{The lowest QP energies in units of $\hbar\omega$ at $T=0$ for the
normal phase 
($\times$) and the superfluid phase (Solid line).}
\label{QPT0}
\end{figure}

We now discuss these results, considering first the response for the normal phase. 
The resonance peaks for $T=0$ and $T'=6$ are relatively narrow 
on the scale of $\omega$. This is perhaps surprising,
as one might expect the Hartree field to wash out the shell-structure 
of the QP spectrum in the normal phase~\cite{BruunN}. 
To understand this, we plot in Fig.\ref{QPT0} 
the lowest QP energies $E_\eta$ for the gas 
in the normal phase at $T=0$. To simplify the plot, we include 
only even values of $l$. The QP energies with odd $l$ behave in a 
completely analogous way. All energies are positive; 
negative normal-phase particle energies are simply holes 
($u_\eta(r)=0$) with positive energy in this representation. For $T=0$, only the
 $\delta(\hbar\tilde{\omega}-E_\eta-E_{\eta'})$
term in Eq.(\ref{Abs}) is non-zero. The thick vertical
arrow in Fig.\ref{QPT0} indicates a typical transition:
creation of a hole with energy $E_h$, and a particle with energy $E_p$,
yielding $\hbar\tilde{\omega}=E_h+E_p\simeq 2.2\hbar\omega$.
The analysis of this normal-phase spectrum is 
basically the same as the one presented in Ref.\cite{BruunN}. 
A key result of the present paper is the finding 
that, although the Hartree field has introduced 
a significant dispersion of the QP energies as a 
function of $l$, the dispersion is almost 
the same for each band. Hence, for $\Delta l=0$, the difference 
of energies between two particle bands (or the sum of
energies of a particle and a hole band, as in Fig.\ref{QPT0}) 
varies much less with $l$ than the energies 
themselves, which results in a relatively narrow 
resonance peak. The resonance for $T=0$ is sharper than for 
$T'=6$.  This is because for $T=0$, only the energy bands 
immediately around $\mu_{\rm F}$ contribute to 
the response due to the Fermi exclusion principle,
whereas for higher $T$, there are transitions between several bands that
yield slightly different transition energies. 

\begin{figure}
\centering
\epsfig{file=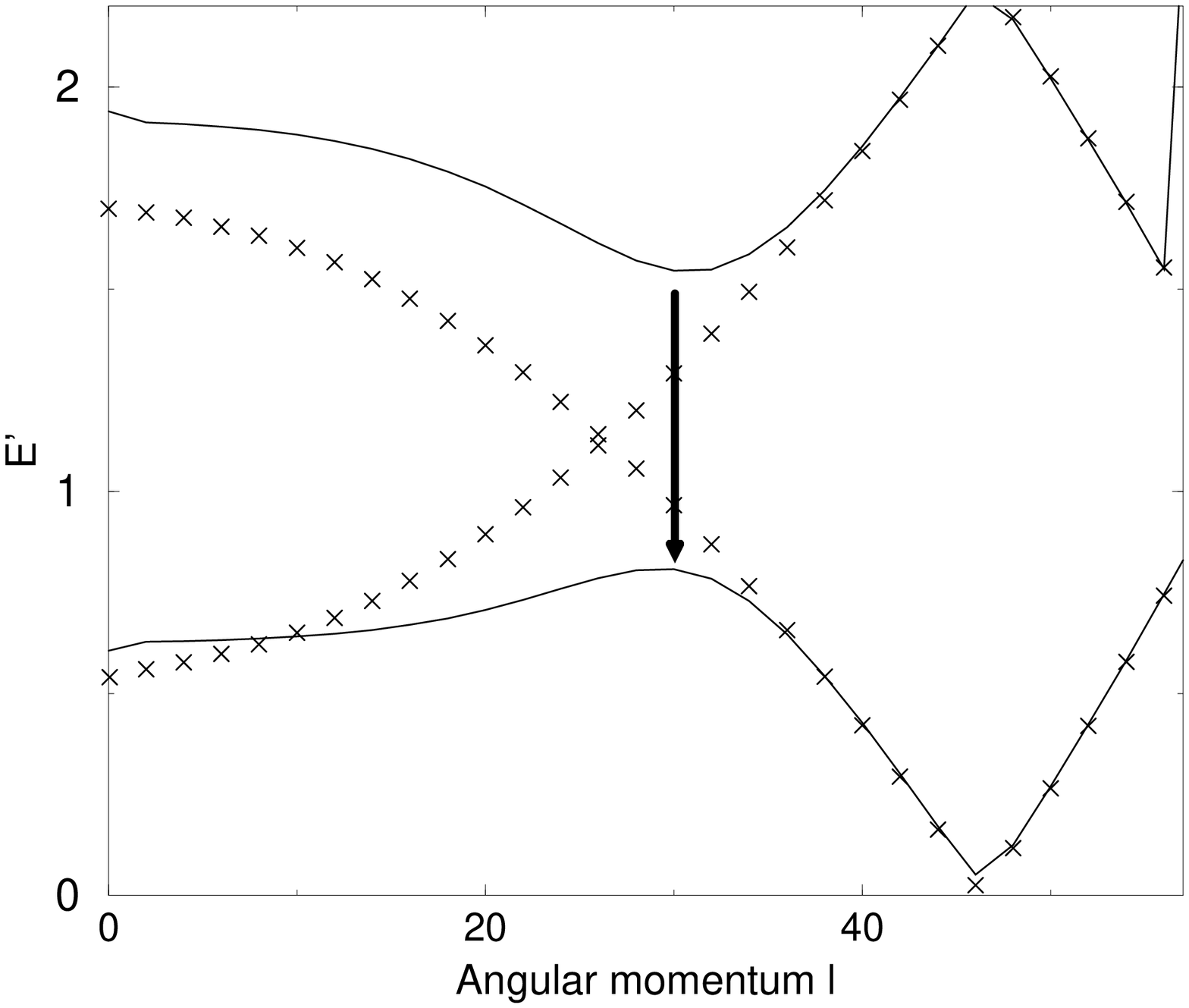,height=0.45\textwidth,width=0.24\textheight,
angle=-90}
\caption{The lowest QP energies in units of $\hbar\omega$ at $T'=4.55$ for
the normal phase 
($\times$) and the superfluid phase (Solid line).}
\label{QPT455}
\end{figure}

We now consider the response when 
the gas is in the superfluid phase.
By comparing the result for $T'=6$ 
and $T'=4.55$ in Fig.\ref{Shakefig}, 
we see that when the gas enters the superfluid phase, 
there is a significant broadening 
of the resonance line. This is due to the fact that
Cooper pairing starts to mix particles with holes,
and the QP spectrum is altered. 
This is depicted in Fig.\ref{QPT455}, 
which shows the lowest even-$l$ QP levels for $T'=4.55$
for both superfluid and normal phases.  
When the energies of particles and holes
are almost degenerate in the normal phase
($l\sim 26$ in Fig.\ref{QPT455}), the pairing 
strongly mixes these two states.  This leads to the usual avoided crossing 
and the QP-spectrum is changed significantly. 
The strong mixing yields the broadening of the resonance line depicted 
in Fig.\ref{Shakefig}. There are now 
transitions with significantly lower
energies than in the normal phase. 
Such a transition, which contributes to the 
$\delta(\hbar\tilde{\omega}+E_{\eta'}-E_\eta)$ 
term in Eq.(\ref{Abs}) with $E_\eta-E_{\eta'}\simeq0.8\hbar\omega$,
is indicated by the vertical arrow in Fig.\ref{QPT455}. 
We also note that the effect of  the pairing decreases with increasing $l$.
This is simply because the centrifugal 
potential ``pushes'' the high $l$ states into the region where the 
order parameter becomes very small. 
The inset in Fig.\ref{Shakefig} shows $\Delta(r)$ and $|W(r)|$. 
For $T'=4.55$, the pairing only takes place around the center of the cloud,
and QP states which have a small amplitude in this region are unaffected. 

For $T\ll T_c$, all the low lying QP states are strongly influenced by the pairing. 
From the inset in Fig.\ref{Shakefig}, we see that 
Cooper pairing now takes place over the entire trapped cloud. The low energy QP spectrum 
for $T=0$ plotted in Fig.\ref{QPT0} is qualitatively 
different from the normal phase spectrum. The low energy QP wave functions 
are centered between the regions where the pairing field and the trapping potential
are significant. These ``in-gap'' states, which were first discussed by 
Baranov~\cite{Baranov}, depend strongly upon the strength of pairing. As $T$ 
decreases and $\Delta(r)$ increases, their energy increases. 
The response of the gas is completely dominated
by these states for $T\ll T_c$. The broad peak for $T=0$ in Fig.\ref{Shakefig} comes from the 
$\delta(\hbar\tilde{\omega}-E_\eta-E_{\eta'})$ 
term in Eq.(\ref{Abs}) with $E_{\eta}=E_{\eta'}$ being the
lowest energy for a given $l$. It reflects excitations of the kind 
${\gamma}^\dagger_{\eta\sigma}{\gamma}^\dagger_{\eta-\sigma}|\Phi_0\rangle$ 
where  $|\Phi_0\rangle$ is the ground state and ${\gamma}^\dagger_{\eta\sigma}$ creates a QP with 
quantum numbers $\eta$ in hyperfine state $\sigma$. 
Hence, for $T\ll T_c$ the response of the gas is a broad peak coming from excitations of 
the lowest QP band. The resonance peak should be 
centered around an increasing frequency as $T$ is lowered, 
since $\Delta(r)$ increases. This is confirmed in Fig.\ref{Shakefig}, 
where a broad peak has emerged in the response for 
 $T'=3.95$, the peak being centered at a lower frequency than for $T=0$.

The qualitative behavior of the response of the gas described above depends on the 
fact that the resonance peaks are relatively well-defined in the normal
phase. We have performed a number of calculations varying both the 
coupling strength and the number of atoms trapped. 
For experimentally realistic parameters, it turns out 
that the Hartree field does not wash out the resonance peaks in the normal phase. 
We therefore believe the analysis above should be valid 
for typical experimental conditions.

\begin{figure}
\centering
\epsfig{file=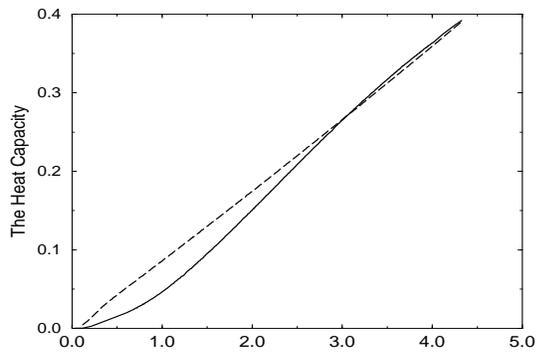,height=0.45\textwidth,width=0.24\textheight,
angle=-90}
\caption{The heat capacity in units of $k_B$ for the normal (dashed) and the
superfluid (solid)
 phase.}
\label{Heatfig}
\end{figure}

The low-$T$ heat capacity is another observable 
which probes the low lying QP spectrum. 
The usual way to measure the energy of a trapped gas 
is to turn off the trapping potential and then deduce the 
velocity distribution from the expanding cloud~\cite{Ensher}. 
As the trapping potential is turned off non-adiabatically, the energy 
observed is really $E_{tot}-E_{pot}$
where $E_{tot}$ is the total energy of the trapped gas and  
$E_{pot}=\sum_\sigma\int\!d^3r \rho_\sigma(r) m\omega^2r^2/2$ 
with $\rho_\sigma(r)=\langle\psi^\dagger_\sigma(r)\psi_\sigma(r)\rangle$. 
Thus, the most appropriate definition of the heat capacity for the 
present purpose is $C_N\equiv\partial_T(E_{tot}-E_{pot})|_{N_\sigma}$. 
Solving the BdG equations with 
varying $T$, we can calculate $E_{tot}(T)$ and $E_{pot}(T)$ and 
therefore $C_N$.  The total energy of the gas given the solution 
of Eq.(\ref{BdG}) is:
\begin{eqnarray}
E_{tot}=\sum_{\eta}f_\eta 
\int d^3r {u_\eta}^*({\mathbf{r}})({\mathcal{H}}_0
+E_\eta)u_\eta({\mathbf{r}})\nonumber\\
+\sum_{\eta}(1-f_\eta)\int d^3r v_\eta({\mathbf{r}})({\mathcal{H}}_0
-E_\eta){v_\eta}^*({\mathbf{r}}).
\end{eqnarray}
As an example, we plot in Fig.\ref{Heatfig}  $c_N\equiv C_N/2N_\sigma$ both 
for the normal and the superfluid phases. 
The number of particles in the trap is held 
constant at $2N_\sigma=24860$  and  $g/(\hbar\omega l_h^3)=-1$, corresponding to a trapping frequency 
of $820$Hz for $^6$Li. The critical temperture for this set of parameters is
$k_BT_c\simeq 4.5\hbar\omega$. From Fig.\ref{Heatfig}, we see that 
the heat capacity is suppressed in the superfluid phase for low $T$. 
This is because the pairing removes the gapless excitations present in the normal phase. 
These excitations have equal particle and hole character, and are therefore 
strongly influenced by the pairing as noted earlier. 
Therefore, the heat capacity is exponentially suppressed by a factor $\sim\exp(\beta\Delta)$, 
where $\Delta$ is the gap in the QP spectrum coming from the Cooper pairing. However, the suppression 
is only significant for $T\ll T_c$, where all angular momentum states are affected 
by the pairing and the QP spectrum is truly gapped (compare Fig.\ref{QPT0} and Fig.\ref{QPT455}). Since the 
system is finite and the superfluid correlations occur gradually starting in the center 
of the trap at $T_c$, and then continuously extending outwards as $T$ decreases
(see inset of Fig.\ref{Shakefig}), there is no discontinuity in the heat capacity at $T=T_c$,
in contrast to the case of an infinite homogeneous system~\cite{Tinkham}. 
It should be noted that it is important that the normal phase 
spectrum is approximately gapless on the scale of $\Delta$ such that the normal phase $c_N$ 
behaves linearly with $T$ for $T<T_c$ as 
depicted in Fig.\ref{Heatfig}. Otherwise, the heat capacity is exponentially suppressed even in 
the normal phase, and one will not observe a significant 
change when the gas becomes superfluid. Fortunately, 
for realistic values of $g$ and $N_\sigma$, it turns 
out that the Hartree field indeed makes the normal phase QP spectrum 
essentially gapless~\cite{BruunN}.

In conclusion, we have presented a detailed analysis of 
two possible ways of detecting the predicted BCS 
phase transition for a trapped gas of fermionic atoms. 
The onset of Cooper pairing 
influences significantly the response of the 
gas to  modulation of the trapping frequency. 
For $T>T_c$, the response has a relatively 
sharp peak, and the width of the peak should narrow as 
$T$ is lowered. Then as  $T=T_c$ is reached,  
one should observe a significant broadening 
of the response peak as the Cooper pairing starts
to affect the low lying QP spectrum. 
For $T\ll T_c$, the low lying QP states are qualitatively 
different from the normal phase states, and the 
response of the gas to the shaking 
is predicted to be a broad peak coming from the 
lowest QP band. The center of the peak 
should move to increasing frequencies  as the pairing 
increases for decreasing $T$. 
Also, one should be able to detect the phase 
transition by looking at the low $T$ heat capacity.
It should be exponentially suppressed for $T\ll T_c$ 
reflecting the gapped nature of the QP 
spectrum due to Cooper pairing. However, a measurement of 
the heat capacity is destructive 
as one has to release the trap and it requires 
several repetitions of the trapping experiment. 
Also, the suppression of $C_{N_\sigma}$ is 
only significant for $T\ll T_c$. We therefore 
estimate that it is a less direct 
way of detecting the transition than by looking at the 
response to a modulation of 
the trapping frequency. The analysis presented 
here should be qualitatively correct for a 
non-spherical symmetric trap as well, although the 
actual calculations would be more cumbersome
 in this case.

\end{document}